\documentclass{article}

\usepackage{times}
\usepackage{bm}
\usepackage{natbib}
\usepackage{graphicx,amsmath,amssymb,amsthm,algorithm,multirow}

\begin{document}




\title{ Efficient Gaussian Process Regression for Large Data Sets}

\author{Anjishnu Banerjee, David Dunson, Surya Tokdar \\
Department of Statistical Science, Duke University, USA \\
email: ab229@stat.duke.edu, dunson@stat.duke.edu, st118@stat.duke.edu}

\maketitle

\newtheorem{theorem}{Theorem}

\begin{abstract}
Gaussian processes (GPs) are widely used in nonparametric regression, classification and spatio-temporal modeling, motivated in part by a rich literature on theoretical properties.  However, a well known drawback of GPs that limits their use is the expensive computation, typically O($n^3$) in performing the necessary matrix inversions with $n$ denoting the number of data points.  In large data sets, data storage and processing also lead to computational bottlenecks and numerical stability of the estimates and predicted values degrades with $n$.  To address these problems, a rich variety of methods have been proposed, with recent options including predictive processes in spatial data analysis and subset of regressors in machine learning.  The underlying idea in these approaches is to use a subset of the data, leading to questions of sensitivity to the subset and limitations in estimating fine scale structure in regions that are not well covered by the subset.  Motivated by the literature on compressive sensing, we propose an alternative random projection of all the data points onto a lower-dimensional subspace.  We demonstrate the superiority of this approach from a theoretical perspective and through the use of simulated and real data examples.\\
Some Keywords: Bayesian; Compressive Sensing; Dimension Reduction; Gaussian Processes; Random Projections; Subset Selection
\end{abstract}

\section{Introduction}  
In many application areas we are interested in modeling an unknown function and predicting its values at unobserved locations. Gaussian processes are used routinely in these scenarios, examples include modeling spatial random effects \citep{BC04,CT92} and supervised classification or prediction in machine learning \citep{R04,S04}. Gaussian processes are mathematically tractable, have desirable properties and provide a probabilistic set-up facilitating statistical inference. 
When we have noisy observations $y_1,\ldots,y_n$ from the unknown function $f:\mathcal{X} \to \Re$ observed at locations $x_1,\ldots,x_n$ respectively, let 
\begin{equation}
\label{e1}
 y_i=f(x_i)+\epsilon_i,\quad \text{for } i =1,\cdots,n,
\end{equation}
where $\epsilon_i$ is the associated idiosyncratic noise. We let $\epsilon_i \sim \mathbb{N}(0,\sigma^2)$ for simplicity.  However, for the techniques we develop here, other noise distributions may be used, including heavy tailed ones. The unknown function $f(\cdot)$ is assumed to be a realization from a Gaussian process with 
mean function $\mu(\cdot)$ and positive definite covariance kernel $k(\cdot,\cdot)$, so that $\mbox{E}\{ f(x) \} = \mu(x)$ and $\mbox{cov}\{ f(x), f(x') \} = k(x,x')$ for all $x,x' \in \mathcal{X}$.  

\par{} The realizations of $f(\cdot)$ at the sample points $x_1,\ldots, x_n$ have a multivariate Gaussian prior, with evaluations of the resulting posterior and computations involved in calculating predictive means and other summaries involving $O(n^3)$ computation unless the covariance has a special structure that can be exploited.  Markov chain Monte Carlo algorithms for posterior computation allowing uncertainty in the residual variance $\sigma^2$ and unknown parameters in the mean function $\mu(\cdot)$ and covariance $k(\cdot)$ may require such computations at every one of a large number of iterations.  Another concern is declining accuracy of the estimates as the dimension increases, as matrix inversion becomes more unstable with the propagation of errors due to finite machine precision. This problem is more acute if the covariance matrix is nearly rank deficient, which is often the case when $f(\cdot)$ is considered at nearby points.

\par{} The above problems necessitate approximation techniques. Most approaches approximate $f(\cdot)$ with another process $g(\cdot)$ that is constrained to a reduced rank subspace.  One popular strategy specifies $g(\cdot)$ as a kernel convolution  \citep{H02}, with related approaches instead relying on other bases such as low rank splines or moving averages \citep{WC99,XG06,KW03}.  A concern with these approaches is the choice of basis. There are also restrictions on the class of covariance kernels admitting such representations.  \citet{BG08} instead proposed a predictive process method that imputes $f(\cdot)$ conditionally on the values at a finite number of knots, with a similar method proposed by \citet{T07} for logistic Gaussian processes.  Subset of regressors \citep{SB01} is a closely related method to the predictive process that was proposed in the machine learning literature and essentially ignored in statistics.  Both of these approaches substantially underestimate predictive variance, with \citet{FB09} proposing a bias correction in the statistics literature and \citet{SG06} independently developing an essentially identical approach in machine learning.  Alternative methods to adjust for underestimation of predictive variance were proposed in \citet{SW03} and \citet{ST03}.  

\par{} \citet{CR05} proposed a unifying framework that encompasses essentially all of these subset of regressors-type approximation techniques, showing that they can be viewed as an approximation to the prior on the unknown function, rather than its posterior.   While these methods do not require choice of a basis, an equally difficult problem arises in determining the location and spacing of knots, with the choice having a substantial impact. In \citet{T07} in the context of density estimation and in unpublished work by Guhaniyogi, Finley, Banerjee and Gelfand in the context of spatial regression, methods are proposed for allowing uncertain numbers and locations of knots in the predictive process using reversible jump and preferential sampling.  Unfortunately, such free knot methods increase the computational burden substantially, partially eliminating the computational savings due to a low rank method.  In the machine learning literature, various optimization methods have been proposed for knot selection, typically under the assumption that the knots correspond to a subset of the data points.  Such methods include online learning \citep{CO03}, greedy posterior maximization \citep{SB01}, maximum information criterion \citep{SW03}, and matching pursuit \citep{KC06} among others. 

In this article, we propose a new type of approximation method that bypasses the discrete knot selection problem using random projections.  The methodology is straightforward to implement in practice, has a theoretical justification and provides a natural generalization of knot-based methods, with pivoted factorizations and the intuitive algorithm of \citet{FB09} arising as special cases.  Motivated by \citet{S06} and \citet{HM09} we use generalized matrix factorizations to improve numerical stability of the estimates, a problem which is typically 
glossed over.  The inspiration for our method arises out of the success of random projection techniques, such as compressed sensing \citep{CT06,D06}, in a rich variety of contexts in machine learning.  Most of this literature focuses on the ability to reconstruct a signal from compressive measurements, with theoretical guarantees provided on the accuracy of a point estimate under sparsity assumptions. In contrast, our goal is to accurately approximate the posterior distribution for the unknown function in a fundamentally different setting.  We also 
explore how these approximations affect inference on the covariance kernel parameters controlling smoothness of the function, an issue essentially ignored in earlier articles. Our theory suggests that predictive process-type approximations may lead to high correlations between the imputed process and parameters, while our method overcomes this problem.
 
\section {Random Projection Approximation Methodology}
\subsection {Predictive Processes and Subset of Regressors}

As a first step, we place the predictive process and subset of regressors methods under a common umbrella.
Consider equation \eqref{e1}, with $\mu \equiv 0$ for notational clarity, and let $X^{*}=\{ x^{*}_1,\ldots, x^{*}_m \}$ denote a set of knots in $\mathcal{X}$. 
Letting $f^{*}=f(X^{*})=\{f(x^{*}_1),\ldots, f(x^{*}_m)\}^{T}$ denote the function $f(\cdot)$ evaluated at the knots, the predictive process
replaces $f(\cdot)$ by $g(\cdot) = E\{ f(\cdot)| f^{*}\}$, with $g(\cdot)$ a kriged surface in spatial statistics terminology \citep{S99}.
 It follows from standard multivariate normal theory that for any $x \in \mathcal{X} \text{,} \, g(x) = (k_{x,*})^{T} (K_{*,*})^{-1}f^{*}$, where $k_{x,*}$ is the $m \times 1$ vector $\{ k(x,x^{*}_1),k(x,x^{*}_2),\ldots, k(x,x^{*}_m)\}^T$ and $K_{*,*}$ is the $m \times m$ matrix with $k(x_i^*,x_j^*)$ in element $i,j$.

Subset of regressors is instead obtained via an approximation to $K_{f,f} =$ cov$\{f(X\})$.  Letting
\[ K_{aug} = \mbox{cov}[ \{ f(X)^T, (f^*)^T \}^T ] =  \left( \begin{array} {cc}
                    K_{f,f} \, \, & K_{f,*} \\
                    K_{*,f} \, \, & K_{*,*}
                   \end{array} \right),
\]
an optimal (in a sense to be described later) approximation to $K_{f,f}$ is obtained as $Q_{f,f}=K_{f,*} (K_{*,*})^{-1} K_{*,f}$, with $Q_{i,j} = K_{i,*}(K_{*,*})^{-1}K_{*,j}$ denoting cell $(i,j)$ of $Q_{f,f}$.This approximation $Q_{f,f}$ is equivalent to $\mbox{cov}\{ g(X)\}$ obtained from the predictive process approximation, and hence the two approaches are equivalent.  
As shown in \citet{CR05}, $g(\cdot)$ is effectively drawn from a Gaussian process with the degenerate covariance kernel
\begin{equation*}
 q_{SOR}(x,z)=(k_{x,*})^{T}(K_{*,*})^{-1}k_{*,z},
\end{equation*}
 where $k_{*,z}=\{ k(x^{*}_1,z),\ldots, k(x^{*}_m,z)\}^T$. From equation \eqref{e1}, we obtain
 $Y = (y_1,\ldots,y_n)^T \sim \mathbb{N}(0; \sigma^2 I + K_{f,f})$ in marginalizing out $f$ over the exact prior $f \sim \mbox{GP}(0, k)$. 
 If we use the approximated version, we have
\begin{equation}
\label{e2}
 y_i = g(x_i) + \epsilon_i,\quad \epsilon_i \sim \mathbb{N}(0,\sigma^2).
\end{equation}
Marginalizing out $g$, we obtain $Y \sim \mathbb{N}(0, \sigma^2 I + Q_{f,f})$.

Let $Y_o$ denote the vector of length $n_o$ of observed values and $Y_p$ the vector of length $n_p$ of values to predict, with $n_o + n_p = n$.  Under the above approximation, the conditional predictive distribution is $(Y_p|Y_o) \sim \mathbb{N}\{Q_{p,o}(Q_{o,o} + \sigma^{2} I)^{-1}Y_o,Q_{o,o}-Q_{p,o}(Q_{o,o} + \sigma^{2}I)^{-1}Q_{o,p}\}$, with $Q_{o,o},Q_{o,p},Q_{p,o}$ denoting submatrices of $Q_{f,f}$.  Using the  Woodbury matrix identity \citep{H08} yields $(Q_{o,o} + \sigma^{2} I)^{-1} = \sigma^{-2}\{ I - K_{o,*}(\sigma^2K_{*,*}+K_{*,o}K_{o,*})^{-1}K_{*,o}\}$, with calculation involving an $m \times m$ matrix. 

\par{}  \citet{FB09} show that the predictive process systematically underestimates variance, since at any $x \in \mathcal{X}$, 
$\mbox{var}\{ f(x) \} - \mbox{var}\{ g(x) \} = \mbox{var}\{ f(x)\, |\, f^{*}\} > 0$. To adjust for this underestimation, they replace $g(\cdot)$ by 
 $g(\cdot)+\epsilon_g(\cdot)$, with $\epsilon_g(x) \sim \mathbb{N}\{ 0,k(x,x)-k_{x,*}^{T} (K_{*,*})^{-1}k_{x,*}\}$ and $\mbox{cov}\{ \epsilon_g(x_1),\epsilon_g(x_2)\} =0$ for 
 $x_1 \neq x_2$.  Hence, in place of equation \eqref{e2}, we have
\begin{equation*}
 y_i = g(x_i) + \epsilon_g(x_i) + \epsilon_i,\quad \epsilon_i \sim \mathbb{N}(0,\sigma^2).
\end{equation*}
A variety of methods for addressing the variance under-estimation problem were independently developed in the machine learning literature \citep{CR05}, with the fully independent training conditional approximation corresponding exactly to the \citet{FB09} approach.  \citet{SG06} also proposed this approach under the sparse Gaussian process with pseudo inputs moniker.  In each of these cases, $g_{M}(\cdot) = g(\cdot) + \epsilon_g(\cdot)$ is effectively drawn from a Gaussian process with the degenerate covariance kernel    
\begin{equation*}
 q_{FITC}(x,z)= q_{SOR}(x,z) +\delta(x,z)\{k(x,z)-q_{SOR}(x,z)\},
\end{equation*}
 where $\delta(x,z) = 1$ if $x=z$ and 0 otherwise.  Our proposed random projection method will generalize these knot-based approaches, leading to some substantial practical advantages.

\subsection{Generalization: Random Projection Method}
The key idea for random projection approximation is to use $g_{RP}(\cdot)=E\{ f(\cdot)|\Phi f(X)\}$ instead of $g(\cdot)=E\{ f(\cdot)|f^*\}$, where $\Phi$ is some $m \times n$ matrix. The approximation $g_{RP}(\cdot)$ is drawn from a Gaussian process with covariance kernel,
\begin{equation*}
 q_{RP}(x,z)=(\Phi k_{x,f})^{T}(\Phi K_{f,f} \Phi^{T})^{-1} \Phi k_{f,z},
\end{equation*}
where $k_{x,f}=\{ k(x,x_1),\ldots, k(x,x_n)\}^T$ and $k_{f,z}=\{ k(x_1,z),\ldots, k(x_n,z)\}^T$. As in the methods of \S$2\cdot 1$, we face the variance under-estimation issue with $\mbox{var}\{ f(x) \} - \mbox{var}\{g_{RP}(x)\} = \mbox{var}\{ f(x)\, |\, \Phi f(X) \} > 0$.  Following the same strategy for bias correction as in \citet{FB09}, we let $g_{RM}(\cdot)$ denote the modified random projection approximation having covariance kernel 
\begin{equation}
\label{e3}
 q_{RM}(x,z)=q_{RP}(x,z) +\delta(x,z)\{ k(x,z)-q_{RP}(x,z)\}.
\end{equation}
When $\Phi$ is the submatrix formed by $m$ rows of a permutation matrix of order $n$ \citep{H08}, we revert back to the formulation of \S$2\cdot 1$, where the knots are an $m$ dimensional subset of the set of data locations $X$.  We consider more generally $\Phi \in  \mathcal{C}$, the class of random matrices with full row-rank and with row-norm $= 1$ to avoid arbitrary scale problems. Before discussing construction of $\Phi$, we consider some properties of the random projection approach.  

\subsection{Properties of the RP method}
\par{(1) Limiting Case:} When $m=n$, $\Phi$ is a square non-singular matrix. Therefore, $(\Phi K_{f,f}\Phi^{T})^{-1} = (\Phi^{T})^{-1}K_{f,f}^{-1}\Phi^{-1}$, so that $Q_{f,f}^{RP}=K_{f,f}$, and we get back the original process with a full rank random projection.
\\

\par{(2) Optimality in terms of Hilbert space projections:} It is well known in the theory of kriging \citep{S99} that taking conditional expectation gives the orthogonal projection into the corresponding space of random variables. Let $\mathcal{H}\{ f(X),\Phi\}$ denote the Hilbert space spanned by linear combinations of the $m$ random variables 
$\Phi f(X)$ and equipped with the inner product  $\left\langle f_1,f_2 \right \rangle = E(f_1 f_2)$ for any $f_1,f_2 \in \mathcal{H}\{ f(X),\Phi \}$.  The orthogonal projection of $f$ to the Hilbert space is $f^{opt}  = argmin_{h \in \mathcal{H}\{f(X,\Phi\}}\|f - h\|$.  From kriging theory 
 $f^{opt}(x) = (\Phi k_{x,f})^{T}(K_{f,f})^{-1} \Phi f(X) = E\{ f(x)|\Phi f(X) \}$.  Hence, the random projection approximation is optimal in this sense.
As $f^{opt}$ is a function of $\Phi \in \mathcal{C}$, the best possible random projection approximation to $f$ could be obtained by choosing $\Phi$ to minimize $\| f^{opt} - f\|$. As the predictive process-type approaches in \S$2\cdot 1$ instead restrict $\Phi$ to a subset of $\mathcal{C}$, the best possible approximation under such approaches is never better than that for the random projection.  While finding the best $\Phi$ is not feasible computationally, \S$3$ proposes a stochastic search algorithm that yields approximations that achieve any desired accuracy level with minimal additional computational complexity.  
\\

\par{(3) Relationship with partial matrix decompositions:} We briefly discussed in \S$2\cdot 1$ that the approximations in the machine learning literature were viewed as reduced rank approximations to the covariance matrices. Here we make an explicit connection between matrix approximation and our random projection scheme, which we build on in the next section.  The Nystr\"om scheme \citep{DM05} considers the rank $m$ approximations to $n \times n$ positive semidefinite matrix $A$ using $m \times n$ matrix $B$, by giving an approximate generalized Choleski decomposition of $A$ as $CC^{T}$, where $C= (BA)^{T}(BAB^{T})^{-1/2}$. The performance of the Nyst\"om scheme depends on how well the range of $B$ approximates the range of $A$. As in property (1), let $Q_{f,f}^{RP}$ be the random projection approximation to $K_{f,f}$. It is easy to see that $Q_{f,f}^{RP}$ corresponds to a Nystr\"om approximation to $K_{f,f}$, with $C=(\Phi K_{f,f})^{T}(\Phi K_{f,f}\Phi^{T})^{-1/2}$.
\par{} The Nystr\"om characterization allows us to obtain a reduced singular value decomposition utilizing the positive definite property as considered in detail in \S$3$. The partial Choleski decompositions for the covariance matrices, advocated in \citet{FW09} for approaches in \S$2\cdot 1$, arise as special cases of the Nystr\"om scheme using permutation submatrices; arguing on the lines of property (2), best case accuracy with the random projection is at least as good as the partial Choleski decomposition. We later show empirically random projection performs substantially better. \\

\par{(4) Relationship with truncated series expansions:} The random projection approximation also arises from a finite basis approximation to the stochastic process $f$. Under the Karhunen-Lo\'eve expansion \citep{A90},
 $$f(x) = \sum_{i=1}^{\infty} \eta_i (\lambda_i)^{1/2} e_i (x), \quad x \in \mathcal{X},$$
where $\mathcal{X}$ is compact and $\lambda_i, e_i$ are eigenvalues and eigenvectors, respectively, of the covariance function $k$, given by the Fredholm equation of the second kind as \citep{G02},
\begin{equation*}
 \int_{\mathcal{X}} k(x_1,x) e_i(x) dx = \lambda_i e_i(x_1), \quad x \in \mathcal{X}.
\end{equation*}
$\eta_i$'s are independent $\mathbb{N}(0,1)$ random variables by virtue of properties of the Gaussian process. Using Mercer's theorem, which is generalization of the spectral theorem for positive definite matrices, we can express the covariance function as \citep{G02},
\begin{equation*}
 k(x_1,x_2) = \sum_{i=1}^{\infty} \lambda_i e_i(x_1) e_i(x_2), \quad x_1,x_2 \in \mathcal{X}.
\end{equation*}
Assume that the eigenvalues in each of the above expansions are in descending order. Let $f_{tr}(x) = \sum_{i=1}^{m} \eta_i (\lambda_i)^{1/2} e_i (x)$ be the approximation to $f(x)$ obtained by finitely truncating the Karhunen-Lo\'eve expansion, keeping only the $m$ largest eigenvalues. The covariance function for $f_{tr}$ is given by $k_{tr}(x_1,x_2) =  \sum_{i=1}^{m} \lambda_i e_i(x_1) e_i(x_2), \quad x_1,x_2 \in \mathcal{X}$, which is, as expected, a corresponding truncation of the expression in Mercer's theorem. If we now evaluate the truncated covariance function on the set of points of interest, $X$, we get the covariance matrix, $K_{tr} = E \Lambda E^{T}$, where $E$ is the $n \times m$ matrix with $(i,j)^{th}$ element given by $e_j(x_i)$ and $\Lambda$ is a  $m \times m$ diagonal matrix with the $m$ eigenvalues in its diagonal. The Karhunen Lo\'eve expansion considers orthogonal functions so that $\int_{\mathcal{X}} e_i(x) e_j(x) dx =0$ whenever  $i \neq j $. If we use the quadrature rule with equal weights for approximation of the integral with the $n$ locations of interest, we have $\sum_{l=1}^{n}e_i(x_l)e_j(x_l) =0$, which means that the matrix $E$ is approximately row-orthogonal. Assuming that $E$ is exactly orthogonal the truncated Mercer expansion matrix $K_{tr}$ is essentially a reduced rank $m$ spectral decomposition for the actual covariance matrix. The covariance matrix of the random vector $g_{RP}(X)$ is equal to the rank $m$ spectral decomposition when we choose the projection matrix $\Phi$ equal to the first $m$ eigenvectors of the actual covariance matrix, as shown in the next section. Therefore $g_{RP}(X)$ has the same probability distribution as $f_{tr}(X)$. In other cases, when $\Phi \neq$ the eigenvectors, as in approaches in section $\S 2\cdot 1$, its easy to show that the random projection corresponds to some other truncated basis expansion in the same way as above. The Karhunen Lo\'eve is however the optimal expansion in the sense that for each $m$, for any other $h_{tr}(\cdot)$ from some $m$ truncated basis expansion, $\int_{\mathcal{X}}E[\{f(x)-h_{tr}(x)\}^{2}] dx$ is minimized over $h_{tr}(\cdot)$, for $h_{tr}(\cdot) = f_{tr}(\cdot)$ \citep{G03}. 

\section{Matrix Approximations \& Projection Construction}
\subsection{Reduced rank matrix approximations}
We introduce stochastic matrix approximation techniques that enable us to calculate nearly optimal projections. We start with some key concepts from linear algebra. Let $\|\cdot\|_2$ and $\|\cdot\|_F$ denote the spectral and Frobenius norm for matrices and let K be any $n \times n$ positive definite matrix. We focus entirely on positive definite matrices. A spectral decomposition  of $K$ is given by, $K = UDU^{T}$, where $D$ is a diagonal matrix whose diagonal elements are the eigenvalues. Since $K$ is positive definite, this is equivalent to the singular value decomposition and the eigenvalues are equal to the singular-values  and $>0$.  $U$ is an orthonormal matrix whose columns are eigenvectors of $K$. Consider any  $ n \times n $ permutation matrix $P$, and since $PP^T=I$ we have, 
\begin{equation*}
 UDU^T = UPP^TDPP^TU^T =  (UP) (PDP^T) (UP)^T.
\end{equation*}
Therefore any permutation of the singular values and their respective vectors leads to an equivalent spectral decomposition for $K$, and it can be shown that the spectral decomposition is unique up to permutations. Henceforth we shall consider only the unique spectral decomposition in which the diagonal elements of $D$ are ordered in increasing order of magnitude,  $d_{11} \geq d_{22} \ldots \geq d_{nn}$. Consider the following partition for the spectral decomposition,
\[ K = \left[ U_m \, U_{(n-m)} \right] \left[ \begin{array} {cc}
                    D_{mm} \,  & 0 \\
                    0 \,  & D_{(n-m)(n-m)}
                   \end{array} \right] \left[ U_m \, U_{(n-m)} \right] ^{T},
\]
where $D_{mm}$ is the diagonal matrix containing the $m$ largest eigenvalues of $K$ and $U_m$ is the $n \times m$ matrix of corresponding eigenvectors. Then it follows from the Eckart-Young theorem \citep{S93} that the best rank m approximation to $K$ is given by $K_m =U_m D_{mm} U_m^T$, in terms of both  $\|\cdot\|_2$ and $\|\cdot\|_F$ . In fact it can be shown that  $\| K-K_m\|_F^2 = \sum_{i=m+1}^{n} d_{ii}^{2}$. 
\par{} Recall that the crux of our random projection scheme was replacing the covariance matrix $K$ by $(\Phi K)^{T}(\Phi K \Phi^{T})^{-1}(\Phi K)$, where $\Phi$ is our random projection matrix. Now if we choose $\Phi = U_m^{T}$, then,
\begin{align*}
 (\Phi K)^{T}(\Phi K \Phi^{T})^{-1}(\Phi K) &= (U_m^{T} K)^{T}(U_m^{T}K U_m)^{-1}(U_m^{T} K)\\
                                       &= \{(D_{mm} 0)U^{T}\}^{T}(D_{mm})^{-1}\{(D_{mm} 0)U^{T}\} = K_m,
\end{align*}
where $0$ above is an $m \times (n-m)$ matrix of zeroes. Therefore the best approximation in our scheme is obtained when we have the first $m$ eigenvectors of the SVD forming our random projection matrix. 
\par{} The problem however is that obtaining the spectral decomposition is as burdensome as computing the matrix inverse, with $O(n^3)$ computations involved. Recent articles in machine learning in the field of matrix approximation and matrix completion have devised random approximation schemes which give near optimal performance with lesser computational cost \citep{HM09,S06}. We can consider these stochastic schemes to address either (i) Given a fixed rank $m$, what is the near optimal projection for that rank and what is the corresponding error; or (ii) Given a fixed accuracy level $1-\epsilon$, what is the near optimal rank for which we can  achieve this and the corresponding projection. We consider each of these questions below.
\par{} We first address the fixed rank problem. For any matrix $K$ of order $n \times n$ and a random vector $\omega$ of order $n \times 1$, $K \omega$ is a vector in the range of $K$.  For an $n \times r$ random matrix $\Omega$ with independent entries from some continuous distribution, $K \Omega$ gives $r$ independent vectors in the range of $K$ with probability $1$. There can be at most $n$ such independent vectors, since the dimension of the range $=n$. As we mentioned earlier, when we evaluate the Gaussian process at a fine grid of points, the covariance matrix $K$ is often severely rank deficient and we should be able to accurately capture its range with $m << n$ vectors. 
\par{} The  next question is how to choose the random matrix $\Omega$. The product $K\Omega$ embeds the matrix $K$ from a $\mathcal{R}^{n\times n}$ space into a $\mathcal{R}^{n \times r}$ space. Embeddings with low distortion properties have been well studied and Johnson-Lindenstrauss transforms \citep{JL86,DG03} are among the most popular low dimensional projections. A matrix $\Omega$ of order $n \times r$ is said to be a Johnson-Lindenstrauss transform for a subspace $V$ of $\mathcal{R}^{n}$ if $| \, \|v \Omega\| - \|v\| \, |$ is small for all $v \in V$ with high probability. For the precise definition of the transform, we refer the readers to Definition 1, \citet{S06}. Initially it was shown that $\Omega$ with $(i,j)^{th}$ element $= (\frac{1}{\surd r} \omega_{ij})$, where $\omega_{ij} \text{ independent} \sim \mathbb{N}(0,1)$ would have Johnson-Lindenstrauss property. Later it has been shown that $\omega_{ij}$'s may be considered to be independent Rademacher or coming from a uniform distribution from the corresponding hypersphere \citep{A03,AV06}. The compressive sensing literature has dealt with these choices in some detail and has found no substantial gain in accuracy in signal compression in using one kind over the other \citep{CT06,D06} - our experiments in the present context concur.
\par{} Having formed $K\Omega$ the concluding step in our matrix approximation scheme is to find $\Phi$. We first perform a low distortion low dimensional Johnson-Lindenstrauss embedding for the covariance matrix and perform the rank $m$ projection for this embedding to come up with $\Phi$. It is easy to then calculate the approximate spectral decomposition of the covariance based on the Nystr\"om approximation for the random projection. The exact steps are shown below in Algorithm \ref{a1} which combines ideas from \citet{S06} and algorithm 5.5 in \citet{HM09}.
\begin{algorithm}
\caption{Approximate spectral decomposition via Nystr\"om method for target rank $m$}
 \label{a1}
 \textit{Given a positive definite matrix K of order $n \times n$  and a randomly generated Johnson-Lindenstrauss matrix $\Omega$ of order $r \times n$, we find the projection matrix $\Phi$ of order $m \times n$ which approximates the range and compute the approximate SVD decomposition via Nystr\"om approximation with $\Phi$}.\\
 1. Form the matrix product $K \Omega$.\\
 2. Compute $\Phi^{T}$ = left factor of the rank $m$ spectral projection of the small matrix $K \Omega$.\\
 3. Form $K_1 = \Phi K \Phi^{T} $ .\\
 4. Perform a Choleski factorization of $K_1 = B B^{T}$.\\
 5. Calculate the Nystr\"om factor $ C = K \Phi^{T} (B^{T})^{-1}$. \\
 6. Compute a spectral decomposition for $C = U D V^{T}$. \\
 7. Calculate the approximate spectral decomposition for $K \approx K_{tr} = U D^{2} U^{T}$.
 
\end{algorithm}
\par{} We give the following result for the approximation accuracy of Algorithm \ref{a1}, which is a modification of theorem 14 in \citet{S06}.
\begin{theorem}
\label{th1}
 Consider any $0 < \epsilon \leq 1$ and $r = \lfloor \frac{m}{\epsilon} \rfloor$. Obtain $K_{tr}$ from Algorithm \ref{a1} for the positive definite matrix $K$ and let $K_m$ be the best rank $m$ approximation for $K$ if terms of $\| \cdot \| _{F}$. Then,\\
\begin{equation*}
\mbox{pr} \{\| K - K_{tr}\| \leq (1 + \epsilon) \| K - K_m \|_F\} \geq \frac{1}{2}
\end{equation*}
\end{theorem}
\par{} With the advances in parallel computing technology and current stress on GPU computing, we may implement a parallel version of Algorithm \ref{a1} by running steps $1 \, \& \, 2$ in parallel for several copies of the matrix $\Omega$; with $log(\frac{1}{\eta})$ copies, we can sharpen the probability in theorem \ref{th1} to $1-\eta$. In our implementations of algorithm \ref{a1} we use $r=m$. The algorithm involves decomposition of the small matrix $\Phi K \Phi^{T}$ which involves $O(m^3)$ operations. The matrix multiplications involved, for example in computing $K_1$ are $O(n^2 m)$, which is the additional cost we pay to have the random projection generalization of the algorithms in \S$2\cdot 1$. Matrix multiplication can be done in parallel, indeed it is the default approach in standard linear algebra packages such as BLAS3 used in Matlab versions 8 and above, and the constants associated with the order of complexity for matrix multiplication is lower than that for inversion. Our results section indicate that added computational complexity in  terms of real CPU time  is indeed negligible for the random projection algorithm versus techniques in \S$2\cdot 1$. In fact with the target error algorithm below, we often achieve lower times than predictive process type approaches of \S$2\cdot 1$, since the rank required to achieve the target error is substantially smaller.
\par{} We now answer the fixed accuracy level question. The eigenvector matrix $U$ from the SVD captures the column space/range of the matrix $K$, in the sense that $K = U U^{T}K$. In general we consider the error in range approximation $\| K - \Phi^{T}\Phi K\|_{\eta}$ ($\eta=2$ or F), as it makes it easier to evaluate the target accuracy. Using simple linear algebra, $U_m U_m^{T} K = K_m$, so that the best rank $m$ range approximator is the same as the rank $m$ SVD approximation. It suffices to then search for good range approximators, since lemma 4 in \citet{DM05} and discussion in \S$5.4$, \citet{HM09} show that the error with the Nystr\"om approximator is at least as small as the error in range approximation, and  empirically is often substantially smaller. We need only find the projection matrix $\Phi$ for the range approximation given the target error level and computation of the approximate spectral decomposition using this $\Phi$ proceeds as in steps $3-7$ of Algorithm \ref{a1}. $\Phi$  can be obtained to satisfy any target error level by trivial modification of steps from algorithm $4.2$ in \citet{HM09} in place of steps $1 \, \& \, 2$ in Algorithm \ref{a1}, summarized below in Algorithm \ref{a2}.
\begin{algorithm}
\caption{Finding range satisfying target error condition}
 \label{a2}
 \textit{Given a positive definite matrix K of order $n \times n$ and target error $\epsilon > 0$ , we find the projection matrix $\Phi$ of order $m \times n$ which gives $\| K - \Phi^{T}\Phi K\|< \epsilon$ with probability $1-\frac{n}{10^{r}}$}.\\
 1. Initialize $j=0$ and $\Phi = []$, the $0 \times n$ empty matrix. \\ 
 2. Draw $r$ random vectors $\omega^{(1)},\ldots,\omega^{(r)}$ each of order $n \times 1$ with independent entries from $\mathbb{N}(0,1)$.\\
 3. Compute $\kappa^{(i)} = K \omega^{(i)}$ for $i=1,\ldots,r$.\\
 4. Is $\max_{i=1,\ldots,r} (\|\kappa^{(j+i)}\|) < \frac{\epsilon \surd \pi}{10 \surd 2}$ ? If yes, step $11$. If no, step $5$. \\
 5. Recompute $j=j+1$ , $\kappa^{(j)} = [I-\{\Phi^{(j-1)}\}^{T}\Phi^{(j-1)}]\kappa^{(j)}$ and $\phi^{(j)}=\frac{\kappa^{(j)}}{\|\kappa^{(j)}\|}$. \\
 6. Set $\Phi^{(j)}= \left[ \begin{array}{c}
 \Phi^{(j-1)}\\
 \ \{\phi^{(j)}\}^{T}\\
\end{array} \right]$.\\
 7. Draw a $n \times 1$ random vector $\omega^{j+r}$ with independent $\mathbb{N}(0,1)$ entries.\\
 8. Compute $\kappa^{(j+r)}= [I-\{\Phi^{(j)}\}^{T}\Phi^{(j)}] K \omega^{(j+r)}$.\\
 9. Recompute $\kappa^{(i)} = \kappa^{(i)} - \phi^{(j)}\left\langle {\phi^{(j)} , \kappa^{(i)} } \right\rangle $ for $i=(j+1),\ldots,(j+r-1)$.\\ 
 10. Back to target error check in step $4$.\\
 11. Output $\Phi = \Phi^{(j)}$.
 \end{algorithm}
\par{} Step $9$ above is not essential, it ensures better stability when $\kappa$ vectors become very small. In our implementations of algorithm \ref{a2} we use an $r$ such that $\frac{n}{10^{r}}=0.1$ to maintain probability of $0.9$ of achieving the error level. The computational requirements of Algorithm \ref{a2} are similar to that of \ref{a1}, for more details we refer the reader to \S$4\cdot 4$ in \citet{HM09}. Posterior fit and prediction in Gaussian process regression usually involves integrating out the Gaussian process, as indicated in \S$4$. We end this subsection with another result which shows that target error in prior covariance matrix approximation governs the error in the marginal distribution of the data, integrating out the Gaussian process. 
\begin{theorem}
\label{th2}
 Let $Y =(y_1,y_2,\ldots,y_n)^T$ be the observed data points and let $\pi_{full} = \int \pi\{Y,f(X)\} dP\{f(X)\}$, $\pi_{RP} = \int \pi\{Y,g_{RP}(X)\}dP\{g_{RP}(X)\}$ their corresponding marginal distributions. If $\|K_{f,f} - Q^{RP}_{f,f}\|_{F} \leq \epsilon$, which is the error in approximation of the covariance matrix, then the Kullback Leibler divergence between the  marginal distributions from the full and approximated Gaussian process,
\begin{equation*}
 KL(\pi_{full},\pi_{RP}) \leq \left \{ n + \left(\frac{n}{\sigma}\right)^{2}\right \} \epsilon
\end{equation*}
\end{theorem}

\subsection{Conditioning numbers and examples}
The full covariance matrix for a smooth Gaussian process tracked at a dense set of locations will be ill-conditioned and nearly rank-deficient in practice, with propagation of rounding off errors due to finite precision arithmetic, the inverses may be highly unstable and severely degrade the quality of the inference. To see this, consider the simple example with covariance  function $k(x,y)=e^{-0.5(x-y)^2}$, evaluated at the points $0.1,0.2$, which gives the covariance matrix,
\[ K = \left( \begin{array} {cc}
                    1.000 \, \, & 0.995 \\
                    0.995 \, \, & 1.000
                   \end{array} \right),
\]
which yields the inverse, 
\[ K^{-1} = \left( \begin{array} {cc}
                    100.5008 \, \, & -99.996 \\
                    -99.996 \, \, & 100.5008
                   \end{array} \right).
\]
Perturbing the covariance kernel slightly to $k(x,y)=e^{-0.75(x-y)^2}$, yields a very similar covariance matrix,
\[ K_{new} = \left( \begin{array} {cc}
                    1.0000 \, \, & 0.9925 \\
                    0.9925 \, \, & 1.0000
                   \end{array} \right),
\]
with $\|K-K_{new}\|_F=0.0035$. However the inverse of the covariance matrix drastically changes to
\[ K^{-1}_{new} = \left( \begin{array} {cc}
                    67.1679 \, \, & -66.6660 \\
                    -66.6660 \, \, & 67.1679
                   \end{array} \right),
\]
with $\|K^{-1}-K_{new}^{-1}\|_F=66.6665$. With such a small change in the magnitude of some elements, we have a huge change in its inverse, which would lead to widely different estimates and predicted values. The problem is obviously much aggravated in large data sets and in Bayesian settings where there the posterior is explored through several rounds of iterations, say in an Gibbs sampling scheme. How well a covariance matrix $K$ is conditioned may be measured by the conditioning number, $\frac{\sigma_l}{\sigma_s}$, where $\sigma_l,\sigma_s$ are its largest and smallest eigenvalues respectively \citep{D83}. Condition numbers are best when they are close to $1$, very large ones indicate numerical instability - in the example above, the condition number of the matrix $K$ is $\approx 400$. Condition number arguments imply that low rank approximations may not only be necessitated by computational considerations but may indeed be desirable for better inference over the full covariance matrix. It therefore makes practical sense to choose amongst two low rank approximations of comparable rank or accuracy, the one that is better conditioned. We now show empirically how condition number is improved greatly with the random projection approximation over the knot based schemes, when considering either a fixed rank or target error approach.
\par{} We first evaluate with respect to the fixed rank question. Consider a similar covariance kernel as above $k(x,y)=e^{-(x-y)^2}$, and evaluate it over a uniform grid of $1000$ points in $[0.1,100]$, and consider the resulting $1000 \times 1000$ covariance matrix $K$. The condition number of $K \approx 1.0652 \times 10^{20}$, which indicates it is severely ill-conditioned. We now apply Algorithm \ref{a1}, with $r=m$, for different choices of the target rank $m$ and calculate the error in terms of the Frobenius and spectral norms, conditioning numbers and the time required. For each choice of $m$, we also consider the approximation as would given by the approaches of \S$2\cdot 1$  in two ways, (1) randomly selecting $m$  grid points out of the $1000$, call this PP1; and (2) selecting the grid points by the partial Choleski factorization with pivoting, as in \citet{TU}, call this PP2, which can be interpreted as a systematic implementation of the suggested approach in \citet{FB09}. Results are summarized in table \ref{t1} for some values of $m$. The random projection approach clearly has better approximation accuracy than the other methods - this becomes more marked with increase in dimension of the approximation. The condition numbers for the random projection scheme are dramatically better than the other 2 approaches, indicating superior numerical stability and reliable estimates. 
\par{} Next we compare with respect to achievement of a target error level. For the random projection approach, we implement Algorithm \ref{a2}. For this comparison it would be useful to know the best possible rank at which the target error would be achieved if we knew the real spectral decomposition. For this purpose we consider matrices of the form $K=EDE^T$, where $E$ is an orthonormal matrix and $D$ is diagonal. The diagonal elements of $D$, which are the eigenvalues of $K$ are chosen to decay at exponential rates, which holds for smooth covariance kernels \citep{FS05}, with $i^{th}$ element $d_{ii}=e^{-i\lambda}$; for the simulations tabulated, we use $\lambda =0.5,0.08,0.04$ respectively. $E$ is filled with independent standard normal entries and then orthonormalized. Algorithm \ref{a2} for random projections, PP1 and PP2 as above are applied to achieve different Frobenius norm error levels $\epsilon$ for different values of matrix order $n$. Results are shown in table \ref{t2}. Clearly random projection achieves the desired target error level with lower ranks for all different values of $\epsilon$ and $n$; also real CPU times required are comparable, in fact the random projection approach has lower time requirements when the rank differences become significant. 
\par{} Lower target ranks, besides the obvious advantages of computational efficiency and stability, imply lesser memory requirements, which is an important consideration when sample size $n$ becomes very large. Time required for matrix norm calculations for checking target error condition for PP1 or PP2 are not counted in the times shown. All times here as well as in following sections, are in seconds and calculated when running the algorithms in Matlab 7.10 version R2010a on a 64bit CentOS 5.5 Linux machine with a 3.33 Ghz dual core processor with 8Gb of random access memory. The random projection benefits from the default parallel implementation of matrix multiplication in Matlab. Lower level implementations of the algorithms, for example C/C++ implementations would require parallel matrix multiplication implementation to achieve similar times. With a GPU implementation with parallel matrix multiplication, random projection approximation can be significantly speeded up.

\section{Parameter Estimation And Illustrations}
\subsection{Bayesian inference for the parameters} 
\par{} An important part of implementing Gaussian process regression is estimation of the unknown parameters of the covariance kernel of the process. Typically the covariance kernel is governed by $2$ parameters, characterizing its range and scale. We shall consider the squared exponential kernel used earlier, $k(x,y)=\frac{1}{\theta_2} e^{-\theta_1 \|x-y\|^2}$ for simplicity, but the techniques herein shall be more generally applicable. $\theta_1$ and $\theta_2$ are the range and inverse scale parameters respectively. We shall use Bayesian techniques for inference here to fully explore the posterior over all possible values of these parameters, also applying the random projection scheme for repeated iterations of Markov chain samplers will allow us to fully demonstrate its power. 
\par{} For Bayesian inference, we have to specify prior distributions for each of the unknown parameters, namely $\theta_1, \theta_2$ and $\sigma^2$, the variance of the idiosyncratic noise in equation \eqref{e1}. In place of \eqref{e1}, using the random projection, we have,
\begin{equation}
 \label{e4}
 y_i = g_{RM}(x_i) + \epsilon_i,  i =1,\ldots,n.
\end{equation}
Using the bias corrected form for the random projection approximation the prior for the unknown function is, $[g_{RM}(X)|\theta_1,\theta_2] \sim \mathbb{N}(0,Q_{f,f}^{RM})$, where $Q_{f,f}^{RM}=Q_{f,f}^{RP}+ D_{M}$, with $D_{M}$ the diagonal matrix as obtained for variance augmentation from equation \eqref{e3}. Letting $\tau = \sigma^{-2}$ and choosing conjugate priors, we let $\tau \sim \mbox{Ga}( a_1,b_1)$, $\theta_2 \sim \mbox{Ga}( a_2,b_2)$ and $\theta_1 \sim \sum_{h=1}^t (1/t) \delta_{c_t}$, denoting a discrete uniform distribution with atoms $\{c_1,\ldots, c_t\}$.  The $\mbox{Ga}(a,b)$ gamma density is parametrized to have mean $a/b$ and variance $a/b^2$. The priors being conditionally conjugate, we can easily derive the full conditional distributions necessary to implement a Gibbs sampling scheme for the quantities of interest as follows,
\begin{align*}
 [g_{RM}(X)| -]  &\sim \mathbb{N}[ \{(Q_{f,f}^{RM})^{-1}+\tau I\}^{-1}Y, \{(Q_{f,f}^{RM})^{-1}+\tau I\}^{-1} ]\\
 [\tau | -] &\sim \mbox{Ga} [ a_1 + \frac{n}{2}, b_1 + \{Y-g_{RM}(X)\}^{T}\{Y-g_{RM}(X)\} ] \\
 [\theta_2 | -] &\sim \mbox{Ga} ( b_2 + f^{T} Q^{-1} f ) \\
 \text{pr} (\theta_1 &= c_i| -) = c | \text{det} Q_{f,f}^{RM} | ^{-\frac{1}{2}} e^{-\frac{1}{2} g_{RM}(X)^{T}(Q_{f,f}^{RM})^{-1} g_{RM}(X)}
\end{align*}
 where  $Q = \theta_2 Q_{f,f}^{RM}$ and  $c$ is a constant such that $\sum_{i=1}^{t} \text{Prob}(\theta_1=c_i|-) =1$.
We can integrate out the Gaussian process $g_{RM}(X)$ from the model to obtain $Y \sim \mathbb{N}(0,\{Q_{f,f}^{RM} + \tau^{-1} I \})$ - this form is useful for prediction and fitting. We show some relevant computational details for the matrix inversion using the Woodbury matrix identity in the appendix.
\par{} For computational efficiency, we pre-compute the random projection matrix for each of the discrete grid points for $\theta_1$ and the corresponding matrix inverse required for the other simulations. Changes in the parameter $\theta_2$ do not affect the eigendirections, hence we do not recompute the projection matrix $\Phi$ and we can compute the new inverse matrix due to a change in $\theta_2$ by just multiplying with the appropriate scalar. Although other prior specifications are extensively discussed in the literature, we have considered simple cases to illustrate the efficacy of our technique. It is observed that inference for the range parameter $\theta_1$ is difficult and Markov chain Monte Carlo schemes tend to have slow mixing due to high correlation between the imputed functional values and the parameter. The random projection approximation appears to take care of this issue in the examples considered here. 
\subsection{Illustrations}
\par{} We first consider a simulated data example where we generate data from functions corresponding to a mixture of Gaussian kernels in $[0,1]$. We consider functions with $3$ different degrees of smoothness - an almost flat one, a moderately wavy one and a highly wavy one. For each of these functions, we consider $10,000$ equi-spaced points  in $[0,1]$ and we add random Gaussian noise to each point - this constitutes our observed data set $Y$. We randomly select $9,000$ points for model fitting and the rest for validation. We now implement random projection with Algorithm \ref{a2} with a couple of different target error levels $(0.1,0.01)$ referred to as RP. We compare it with predictive process with equispaced selection of knots and with the modified version of knot selection by pivoted Choleski factorization \citep{TU} explained in \S$3.2$, referred to as PP1 and PP2 respectively. In this simulated example, as well as in the real data examples we use the squared exponential covariance kernel with prior specifications as in the previous section. For the idiosyncratic noise, we use hyperparameters $a_1,b_1$ such that the mean is approximately equal to estimated noise precision with ordinary least square regression. In particular for the smooth one we use $a_1=1,b_1=10$. Hyperparameter choices for covariance kernel parameters are guided by some trial runs, we use a grid of $2000$ equispaced points in $[0,2]$ for $\theta_1$ and $a_2=2,b_2=20$ for $\theta_2$.  We run Gibbs samplers for $10,000$ iterations with  the first $500$ discarded for burn-in. We calculate the predicted values for the held-out set with the posterior means of the parameters from the Gibbs iterations and we also calculate the average rank required to achieve the target accuracy over the iterations. Effective sample size is calculated by using the output for the Markov chains with the CODA package in R. The results are tabulated in table \ref{t3}, whereby random projection has substantial gain in predictive accuracy and in the target rank required, as well as substantially better effective sample sizes for the unknown parameters of the covariance kernel as well as for the predicted points. With the predictive process type approaches, we would need substantially more Markov chain Monte Carlo iterations to achieve similar effective sample sizes, leading to an increased computational cost.
\par{} We finally consider a couple of real data examples, which have been used earlier for reduced rank approaches in Gaussian process regression, of contrasting sizes. The first is the abalone dataset, from the UCI machine learning database \citep{FA10}, where the interest is in modeling the age of abalone, given other attributes, which are thought to be non-linearly related to age. The dataset consists of $4000$ training and $177$ test cases. We use Euclidean distance between the attributes for our covariance function for the Gaussian process and for the gender attribute, (male/female/infant) is mapped to $\{ (1,0,0),(0,1,0), (0,0,1)\}$. The other example we consider is the Sarcos Robot arm, where we are interested in the torque as given by the $22^{nd}$ column given the other measurements in the  remaining $21$  columns. This dataset has $44,484$ training and $4,449$ test cases. We once again consider Euclidean distances between the attributes. For each of the experiments, we use Algorithm \ref{a2} with target error level $0.01$. The hyperparameters for each example is chosen in similar fashion as the simulated example. This leads to choosing $a_1=1,b_1=0.1$ for the abalone dataset and $a_1=2,b_1=0.1$ for the Sarcos Robot arm. The grid for $\theta_1$ in either case is $2000$ equispaced points in $[0,2]$; for $\theta_2$, in abalone we have $a_2=1,b_2=1$ while we have $a_2=1,b_2=0.75$ for Sarcos Robot arm.  The Gibbs sampler for the abalone data set is run for $10,000$ iterations with $1,000$ discarded for burn-in, while for the Sarcos Robot arm, it is run for $2,000$ iterations with $500$ discarded for burn-in. 
\par{} The results for both these experiments, tabulated in table \ref{t4}. There is improvement in predictive accuracy when using random projections in both the examples,  in particular for the Robot arm dataset predictive accuracy is significantly better. This improvement perhaps is a consequence of the fact that we get better estimation for the parameters when using the random projection approach. In the Sarcos Robot arm data, both the covariance kernel parameters are readily observed to have different posteriors with the random projection approach. This is a consequence of the poor behavior of the Markov chains for these parameters, they exhibit poor mixing. In fact this problem of poor mixing when approximating a stochastic process by imputed points is not unique to Gaussian processes, it have been observed in other contexts too, possibly due to the chains for the imputed points and the unknown parameters being highly correlated with each other \citep{GW06}. The random projection approach appears to improve this to a great extent by not considering specific imputed points. The inference is not very sensitive to the choice of hyperparameters, with datasets of this size we are able to overcome prior influence if any. In particular in trial runs with smaller number of iterations, changing the grid for $\theta_1$ to $1000$ uniformly spaced points in $[0,1]$ yielded almost similar results, random projection performing better than the knot based approaches.
\section{Concluding Remarks}
\par{} We have developed a broad framework for reduced rank approximations under which almost the entire gamut of existing approximations can be brought in. We have tried to stochastically find the best solution under this broad framework, thereby leading to gains in performance and stability over the existing approaches. Another important contribution has been to connect not only the machine learning and statistical approaches for Gaussian process approximation, but also to relate them to matrix approximations themes - we have shown that the reduced rank Gaussian process schemes are effectively different flavors of approximating the covariance matrix arising therein. The random projection approach has been mainly studied as an approximation scheme in this article, it is also worthwhile considering it from a model based perspective and investigate the added flexibility it offers as an alternative model. 
\par{} We have not explored the performance of parallel computing techniques in this context, though we have indicated how to go about parallel versions of the algorithms at hand. Further blocking techniques and parallelization remains an area of future interest. We also plan on working out the multivariate version of random projection approximations. In ongoing work, we explore similar approaches in other different contexts - a couple of examples being in the context of functional modeling, where the domain may be discrete and also in the case of parameter estimation for diffusion processes - where similar dimensionality problems are faced sometimes in terms of their discrete Euler approximations. In other ongoing work, we also explore the theoretical rates of convergence of the truncated expressions for different classes of covariance kernels and convergence of the associated posterior distributions of the unknown parameters.
\section*{Acknowledgement}
This work was supported by a grant from the National Institute of Environmental Health Sciences.  The content is solely the responsibility of the authors and does not necessarily represent the official views of the National Institute of Environmental Health Sciences or the National Institutes of Health.
\section*{Appendix}
\subsection*{Proof of Theorem 1}
By construction, 
\begin{align*}
K_{tr} &= U D^{2} U^{T} = U D V^{T} V D U^{T} = C C^{T}\\
           & = K \Phi^{T} (B^{T})^{-1} B^{-1} \Phi K = K \Phi^{T} (B B^{T})^{-1} \Phi K \\ 
           & =  K \Phi^{T} K_{1}^{-1} \Phi K = (\Phi K)^{T} (\Phi K \Phi^{T})^{-1} \Phi K 
\end{align*}
This shows that the reduced SVD form, $K_{tr}$ produced by Algorithm \ref{a1} is indeed equal to the random projection approximation, which is equal to a generalized projection matrix as explained below. 
\par{}
The generalized rank $m$ projection matrix for the projection whose range is spanned by the columns of an $n \times m$ matrix $A$, with $m \leq n$ and whose nullity is the orthogonal complement of the range of $n \times m$ matrix B, is given by $A (B^{T}A)^{-1} B^{T}$. This is a generalization of the standard projection matrix formula \citep{D91}.  Therefore, $K_{tr}=PK$, where $P= K \Phi^{T} \{\Phi (K \Phi^{T})\}^{-1} \Phi$ is the generalized projection matrix with range spanned by the columns of $K\Phi^{T}$ and whose nullity is the orthogonal complement of the range of $\Phi^{T}$. Again, by construction, range of $\Phi^{T}$ = range of $K\Omega$ and therefore, range of $K \Phi^{T}=$ range of $K^{2}\Omega =$ range of $K\Omega$. Finally since range of $K \Omega$ = row-space of $\Omega^{T}K$, the result follows by a direct application of theorem $14$ in \citet{S06}.
\subsection*{Proof of Theorem 2}
The Kullback Leibler divergence between two $n-$variate normal distributions $\mathcal{N}_0 = \mathbb{N}(\mu_0,\Sigma_0) \,\, \& \,\, \mathcal{N}_1 = \mathbb{N}(\mu_1,\Sigma_1)$ is given by,
\begin{equation*}
\text{KL}(\mathcal{N}_0 \| \mathcal{N}_1) = \frac{1}{2} \left[ \mathrm{tr} \left( \Sigma_1^{-1} \Sigma_0 \right)  - n -\log \left\{ { \det \left (\Sigma_1^{-1}\Sigma_0 \right )  } \right\} + \left( \mu_1 - \mu_0\right)^{T} \Sigma_1^{-1} ( \mu_1 - \mu_0 )  \right ]
\end{equation*}
In our case, $\mathcal{N}_0 = \pi_{full} = \mathbb{N}(y; 0, K_{f,f} + \sigma^2 I)$ and $\mathcal{N}_1 = \pi_{RP} = \mathbb{N}(y; 0, Q^{RP}_{f,f}+\sigma^2 I)$. Therefore $KL(\pi_{full},\pi_{RP}) = \frac{1}{2} \left[ \mathrm{tr} \left( \Sigma_1^{-1} \Sigma_0 \right)  - n -\log \left\{ { \det \left (\Sigma_1^{-1}\Sigma_0 \right )  } \right\} \right ]$, with $\Sigma_0 = K_{f,f} + \sigma^2 I $ and $\Sigma_1 = Q^{RP}_{f,f} + \sigma^2 I$. We have $\|\Sigma_0-\Sigma_1\|_{F} = \|K_{f,f} - Q^{RP}_{f,f}\|_{F} \leq \epsilon$.
\par{} Break the expression for the Kullback Leibler divergence into $2$ parts with the first part,\\ $\mathrm{tr} \left ( \Sigma_1^{-1} \Sigma_0 \right)- n =  \mathrm{tr} \left\{ \Sigma_1^{-1}(\Sigma_0 -\Sigma_1) \right\}= \sum_{i=1}^{n}\sum_{j=1}^{n} s_{ij}d_{ji}$ where $s_{ij},d_{ji}$ are $(ij)^{th}\, \& \, (ji)^{th}$ elements of $\Sigma_1^{-1},(\Sigma_0 -\Sigma_1)$ respectively. Then,
\begin{equation}
\label{e5}
 \mathrm{tr} \left ( \Sigma_1^{-1} \Sigma_0 \right)- n \leq \|\Sigma_1^{-1}\|_{max} \sum_{i=1}^{n}\sum_{j=1}^{n} d_{ji}
                                                       \leq \|\Sigma_1^{-1}\|_{max} \, n^{2} \epsilon
\end{equation}
In the inequality above we use  $\|\Sigma_1^{-1}\|_{max} = \max_{ij} s_{ij}$ and the fact that $\|\Sigma_0-\Sigma_1\|_{F} \leq \epsilon \implies \sum_{i=1}^{n}\sum_{j=1}^{n} d_{ji} \leq n^2 \epsilon$. Now $\|\Sigma_1^{-1}\|_{max} \leq \|\Sigma_1^{-1} \|_2$. Since $\Sigma_1^{-1}$ is symmetric postive definite, $\|\Sigma_1^{-1} \|_2$ is the largest eigenvalue of $\Sigma_1^{-1}$ which is equal to the inverse of the smallest eigenvalue of $\Sigma_1$. Recall that $\Sigma_1 = Q^{RP}_{f,f} + \sigma^2 I$ and $Q^{RP}_{f,f}$  is positive semi-definite and has non negative eigenvalues. Therefore all eigenvalues of $\Sigma_1 \geq \sigma^{2}$, and using this in conjunction with inequality \eqref{e5}, we have,
\begin{equation}
\label{e6}
 \mathrm{tr} \left( \Sigma_1^{-1} \Sigma_0 \right)- N \leq \left(\frac{n}{\sigma}\right)^{2} \epsilon
\end{equation}
\par{} It remains to bound the second part of the divergence expression. We have $\det \left (\Sigma_1^{-1}\Sigma_0 \right )  = \left ( \prod_{i=1}^{n} \lambda^{0}_{i} \right ) / \left ( \prod_{i=1}^{n} \lambda^{1}_{i} \right )$, where $\lambda^{0}_i,\lambda^{1}_i$ are eigenvalues of $\Sigma_0 \, \& \, \Sigma_1$ respectively. Since $\Sigma_0, \Sigma_1$ are symmetric, by the Hoffman-Weilandt inequality \citep{B97}, there exists a permutation $p$ such that $\sum_{i=1}^{n}\left \{ \lambda^{0}_{p(i)} -\lambda^{1}_{i} \right \}^{2} \leq \|\Sigma_0-\Sigma_1\|^{2}_{F} \leq \epsilon^2$. Therefore with the same permutation $p$, we have for each $i, \, \left\{ \lambda^{0}_{p(i)} / \lambda^{1}_{i} \right\} \in [1-\epsilon, 1 + \epsilon] $. Trivial manipulation then yields, $ \log \left\{ { \det \left (\Sigma_1^{-1}\Sigma_0 \right )  } \right\}  \in [n \log (1-\epsilon), n \log (1 + \epsilon)]$, so that, 
\begin{equation}
\label{e7}
 -\log \left\{ { \det \left (\Sigma_1^{-1}\Sigma_0 \right )  } \right\}  \leq n \epsilon
\end{equation}
\par{} Combining inequalities \eqref{e6} and \eqref{e7}, we have,
\begin{equation*}
  \text{KL}(\pi_{full},\pi_{RP}) \leq \left \{ n + \left(\frac{n}{\sigma}\right)^{2}\right \} \epsilon
\end{equation*}
which completes the proof.
\par{} This is not an optimal bound, but serves our basic goal of showing that the Kullback Leibler divergence is of the same order as the error in estimation of the covariance matrix in terms of Frobenius norm. Additional assumptions on the eigenspace of the covariance matrix would yield tighter bounds.  
\subsection*{Example of inversion with the Woodbury matrix identity}
Either of the algorithms, \ref{a1} or \ref{a2}, in this paper yield $Q^{RP}_{f,f}= UD^{2} U^{T}$, with $ U^{T}U = I$. We would be interested in calculating $\Sigma_1^{-1} = (Q^{RP}_{f,f} + \sigma^2 I)^{-1}$, in the marginalized form for inference or prediction. Using the Woodbury matrix identity \citep{H08} we have,
\begin{align*}
 \Sigma_1^{-1} &= \sigma^{-2}I - \sigma^{-2}U (D^{-2}+  \sigma^{-2} U^{T}U)^{-1}U^{T} \sigma^{-2}\\
               &= \sigma^{-2}I - \sigma^{-4}U (D^{-2}+  \sigma^{-2} I)^{-1}U^{T}
\end{align*}
In the above $D^{-2}+  \sigma^{-2} I$ is a diagonal matrix whose inverse can be obtained by just taking reciprocals of the diagonals. Thus direct matrix inversion is entirely avoided with the decomposition available from the algorithms.
\bibliographystyle{biometrika} 
\bibliography{mybib} 
\begin{table}[t]
\begin{center}
\begin{tabular}{lllll}
\hline
\multicolumn{1}{c}{For $m=10$}  &\multicolumn{1}{c}{$\| \|_F$} &\multicolumn{1}{c}{$\| \|_2$} &\multicolumn{1}{c}{Cond No} &\multicolumn{1}{c}{Time}
\\ \hline 
RP  & 106.1377 & 17.6578 & 1.0556 & 0.06 \\
PP1         & 107.6423 & 17.6776 & 1.2356  & 0.04\\
PP2         & 106.6644 & 17.6778 & 1.2619  & 0.04\\
\\ \hline 
\multicolumn{1}{c}{For $m=25$}  &\multicolumn{1}{c}{$\| \|_F$} &\multicolumn{1}{c}{$\| \|_2$} &\multicolumn{1}{c}{Cond No} &\multicolumn{1}{c}{Time}
\\ \hline 
RP  & 82.1550 & 17.2420 & 1.7902  & 0.22\\
PP1         & 91.1016 & 17.5460 & 230.236  & 0.18\\
PP2         & 85.6616 & 17.3800 & 13.8971  & 0.21\\
\\ \hline 
\multicolumn{1}{c}{For $m=50$}  &\multicolumn{1}{c}{$\| \|_F$} &\multicolumn{1}{c}{$\| \|_2$} &\multicolumn{1}{c}{Cond No} &\multicolumn{1}{c}{Time}
\\ \hline 
RP  & 50.5356 & 14.2998 & 2.9338 & 0.27\\
PP1         & 79.1030 & 17.0172 & 2803.5  & 0.24\\
PP2         & 69.5681 & 15.6815 & 876.23 & 0.25\\
\\ \hline 
\multicolumn{1}{c}{For $m=100$}  &\multicolumn{1}{c}{$\| \|_F$} &\multicolumn{1}{c}{$\| \|_2$} &\multicolumn{1}{c}{Cond No} &\multicolumn{1}{c}{Time}
\\ \hline 
RP (Algo1)  & 6.6119 & 2.8383 & 20.6504 & 0.40 \\
PP1         & 39.9642 & 13.1961 & 1.3815 $\times 10^{6}$ & 0.31 \\
PP2         & 10.1639 & 6.3082 & 1792.1 & 0.36\\
\\ \hline 
\end{tabular}
\end{center}
\caption{Comparative performance of the approximations in terms of matrix error norms, with the random projection approach based on Algorithm \ref{a1}.}
\label{t1}
\end{table}
\begin{table}[t]
\makebox[0.75\textwidth]{
\begin{tabular}{cc|c|c|c|l}
\cline{3-6}
& & PP1 & PP2 & RP  \\ \cline{1-5}
\multicolumn{1}{|c|}{\multirow{2}{*}{$n=100,\epsilon=0.1,\text{optimal m = 5}$}} &
\multicolumn{1}{|c|}{Required Rank} & 17 & 9 & 7     \\ \cline{2-5}
\multicolumn{1}{|c|}{}                        &
\multicolumn{1}{|c|}{Cond No} & 298.10 & 54.59 & 20.08     \\ \cline{2-5}
\multicolumn{1}{|c|}{}                        &
\multicolumn{1}{|c|}{Time} & 0.03 & 0.04 & 0.07      \\ \cline{1-5}
\multicolumn{1}{|c|}{\multirow{2}{*}{$n=1000,\epsilon=0.01,\text{optimal m = 69}$}} &
\multicolumn{1}{|c|}{Required Rank} & 213 & 97 & 78   \\ \cline{2-5}
\multicolumn{1}{|c|}{}                        &
\multicolumn{1}{|c|}{Cond No} & $2.30\times 10^7$ & 2164.6 & 473.43     \\ \cline{2-5}
\multicolumn{1}{|c|}{}                        &
\multicolumn{1}{|c|}{Time} & 12.1 & 11.5 & 36.2   \\ \cline{1-5}
\multicolumn{1}{|c|}{\multirow{2}{*}{$n=10000,\epsilon=0.01,\text{optimal m = 137}$}} &
\multicolumn{1}{|c|}{Required Rank} & 1757 & 793 &  174  \\ \cline{2-5}
\multicolumn{1}{|c|}{}                        &
\multicolumn{1}{|c|}{Cond No} & $3.19\times 10^{19}$ & $2.30\times 10^9$ & 1012.3     \\ \cline{2-5}
\multicolumn{1}{|c|}{}                        &
\multicolumn{1}{|c|}{Time} & 335 & 286 & 214   \\ \cline{1-5}
\end{tabular}
}
\caption{Comparison of the ranks required to achieve specific target errors by the different algorithms, with random projection based on Algorithm \ref{a2}}
\label{t2}
\end{table}
\begin{table}[t]
\makebox[0.75\textwidth]{
\begin{tabular}{cc|c|c|c|l}
\cline{3-6}
& & PP1 & PP2 & RP  \\ \cline{1-5}
\multicolumn{1}{|c|}{\multirow{2}{*}{$\epsilon=0.1$, smooth}} &
\multicolumn{1}{|c|}{MSPE} & 11.985 & 8.447 & 3.643     \\ \cline{2-5}
\multicolumn{1}{|c|}{}                       &
\multicolumn{1}{|c|}{Avg Required Rank} & 1715.6 & 453.8 & 117.2     \\ \cline{2-5}
\multicolumn{1}{|c|}{}                        &
\multicolumn{1}{|c|}{$95\%$ Interval Required Rank} & [1331,2542] & [377,525] & [97,141]     \\ \cline{2-5}
\multicolumn{1}{|c|}{}                        &
\multicolumn{1}{|c|}{Posterior Mean, $\theta_1$} & 0.09 & 0.10 & 0.06     \\ \cline{2-5}
\multicolumn{1}{|c|}{}                        &
\multicolumn{1}{|c|}{$95\%$ credible interval, $\theta_1$} & [0.05,0.14] & [0.05,0.15] & [0.04,0.08]     \\ \cline{2-5}
\multicolumn{1}{|c|}{}                        &
\multicolumn{1}{|c|}{ESS, $\theta_1$} & 496 & 870 & 1949     \\ \cline{2-5}
\multicolumn{1}{|c|}{}                        &
\multicolumn{1}{|c|}{Posterior Mean, $\theta_2$} & 0.91 & 1.15 & 1.25     \\ \cline{2-5}
\multicolumn{1}{|c|}{}                        &
\multicolumn{1}{|c|}{$95\%$ credible interval, $\theta_2$} & [0.58,1.58] & [0.85,1.43] & [1.09,1.46]     \\ \cline{2-5}
\multicolumn{1}{|c|}{}                        &
\multicolumn{1}{|c|}{ESS, $\theta_2$} & 2941 & 3922 & 4518     \\ \cline{2-5}
\multicolumn{1}{|c|}{}                        &
\multicolumn{1}{|c|}{Avg ESS, Predicted Values} & 2190 & 3131 & 5377     \\ \cline{2-5}
\multicolumn{1}{|c|}{}                        &
\multicolumn{1}{|c|}{Time} & 39761  & 29355 & 32365      \\ \cline{1-5}
\multicolumn{1}{|c|}{\multirow{2}{*}{$\epsilon=0.01$, wavy}} &
\multicolumn{1}{|c|}{MSPE} & 10.114 & 6.891 & 2.265   \\ \cline{2-5}
\multicolumn{1}{|c|}{}                        &
\multicolumn{1}{|c|}{Avg Required Rank} & 3927.1 & 941.5 & 129.7   \\ \cline{2-5}
\multicolumn{1}{|c|}{}                        &
\multicolumn{1}{|c|}{$95\%$ Interval Required Rank} & [2351,5739] & [868,1165] & [103,159]     \\ \cline{2-5}
\multicolumn{1}{|c|}{}                        &
\multicolumn{1}{|c|}{Posterior Mean, $\theta_1$} & 0.07 & 0.08 & 0.13     \\ \cline{2-5}
\multicolumn{1}{|c|}{}                        &
\multicolumn{1}{|c|}{$95\%$ credible interval, $\theta_1$} & [0.01,0.14] & [0.02,0.15] & [0.09,0.17]     \\ \cline{2-5}
\multicolumn{1}{|c|}{}                        &
\multicolumn{1}{|c|}{ESS, $\theta_1$} & 574 & 631 & 1918      \\ \cline{2-5}
\multicolumn{1}{|c|}{}                        &
\multicolumn{1}{|c|}{Posterior Mean, $\theta_2$} & 0.83 & 0.85 & 0.79     \\ \cline{2-5}
\multicolumn{1}{|c|}{}                        &
\multicolumn{1}{|c|}{$95\%$ credible interval, $\theta_2$} & [0.21,1.74] & [0.40.1.63] & [0.45,1.29]     \\ \cline{2-5}
\multicolumn{1}{|c|}{}                        &
\multicolumn{1}{|c|}{ESS, $\theta_2$} & 3679 & 4819 & 5002     \\ \cline{2-5}
\multicolumn{1}{|c|}{}                        &
\multicolumn{1}{|c|}{Avg ESS, Predicted Values} & 2875 & 3781 & 5769   \\ \cline{2-5}
\multicolumn{1}{|c|}{}                        &
\multicolumn{1}{|c|}{Time} & 78812  & 47642 & 33799   \\ \cline{1-5}
\multicolumn{1}{|c|}{\multirow{2}{*}{$\epsilon=0.01$, very wavy}} &
\multicolumn{1}{|c|}{MSPE} & 17.41 & 13.82 & 6.93   \\ \cline{2-5}
\multicolumn{1}{|c|}{}                        &
\multicolumn{1}{|c|}{Avg Required Rank} & 4758.5 & 1412.5 & 404.5   \\ \cline{2-5}
\multicolumn{1}{|c|}{}                        &
\multicolumn{1}{|c|}{$95\%$ Interval Required Rank} & [2871,6781] & [1247,1672] & [312,475]     \\ \cline{2-5}
\multicolumn{1}{|c|}{}                        &
\multicolumn{1}{|c|}{Posterior Mean, $\theta_1$} & 0.11 & 0.09 & 0.05     \\ \cline{2-5}
\multicolumn{1}{|c|}{}                        &
\multicolumn{1}{|c|}{$95\%$ credible interval, $\theta_1$} & [0.04,0.17] & [0.05,0.13] & [0.03,0.08]     \\ \cline{2-5}
\multicolumn{1}{|c|}{}                        &
\multicolumn{1}{|c|}{ESS, $\theta_1$} & 741 & 747 & 1049      \\ \cline{2-5}
\multicolumn{1}{|c|}{}                        &
\multicolumn{1}{|c|}{Posterior Mean, $\theta_2$} & 1.27 & 1.18 & 1.19     \\ \cline{2-5}
\multicolumn{1}{|c|}{}                        &
\multicolumn{1}{|c|}{$95\%$ credible interval, $\theta_2$} & [1.08,1.43] & [1.12,1.41] & [1.15,1.34]     \\ \cline{2-5}
\multicolumn{1}{|c|}{}                        &
\multicolumn{1}{|c|}{ESS, $\theta_2$} & 1521 & 2410 & 2651     \\ \cline{2-5}
\multicolumn{1}{|c|}{}                        &
\multicolumn{1}{|c|}{Avg ESS, Predicted Values} & 1263 & 1415 & 2422   \\ \cline{2-5}
\multicolumn{1}{|c|}{}                        &
\multicolumn{1}{|c|}{Time} & 89715  & 57812 & 47261   \\ \cline{1-5}
\end{tabular}
}
\caption{Simulated data sets with the target error algorithm for the three different simulations. Different algorithms compared in terms of preditive MSE and various posterior summaries for the unknown parameters. ESS stands for effective sample size.}
\label{t3}
\end{table}

\begin{table}
\makebox[0.65\textwidth]{
\begin{tabular}{cc|c|c|c|l}
\cline{3-6}
& & PP1 & PP2 & RP  \\ \cline{1-5}
\multicolumn{1}{|c|}{\multirow{2}{*}{Abalone Dataset}} &
\multicolumn{1}{|c|}{MSPE} & 1.785 & 1.517 & 1.182     \\ \cline{2-5}
\multicolumn{1}{|c|}{}                       &
\multicolumn{1}{|c|}{Avg Reqd Rank} & 417.6 & 328.8 & 57.2     \\ \cline{2-5}
\multicolumn{1}{|c|}{}                        &
\multicolumn{1}{|c|}{$95\%$ Interval Required Rank} & [213,750] & [207,651] & [43,71]     \\ \cline{2-5}
\multicolumn{1}{|c|}{}                        &
\multicolumn{1}{|c|}{Posterior Mean, $\theta_1$} & 0.212 & 0.187 & 0.149     \\ \cline{2-5}
\multicolumn{1}{|c|}{}                        &
\multicolumn{1}{|c|}{$95\%$ credible interval, $\theta_1$} & [0.112,0.317] & [0.109,0.296] & [0.105,0.207]     \\ \cline{2-5}
\multicolumn{1}{|c|}{}                        &
\multicolumn{1}{|c|}{ESS, $\theta_1$} & 516 & 715 & 1543      \\ \cline{2-5}
\multicolumn{1}{|c|}{}                        &
\multicolumn{1}{|c|}{Posterior Mean, $\theta_2$} & 0.981 & 1.014 & 1.105     \\ \cline{2-5}
\multicolumn{1}{|c|}{}                        &
\multicolumn{1}{|c|}{$95\%$ credible interval, $\theta_2$} & [0.351,1.717] & [0.447,1.863] & [0.638,1.759]     \\ \cline{2-5}
\multicolumn{1}{|c|}{}                        &
\multicolumn{1}{|c|}{ESS, $\theta_2$} & 1352 & 1427 & 1599     \\ \cline{2-5}
\multicolumn{1}{|c|}{}                        &
\multicolumn{1}{|c|}{Time} & 19468  & 21355 & 15423       \\ \cline{1-5}
\multicolumn{1}{|c|}{\multirow{2}{*}{Sarcos Robot Arm}} &
\multicolumn{1}{|c|}{MSPE} & 0.5168 & 0.2357 & 0.0471   \\ \cline{2-5}
\multicolumn{1}{|c|}{}                        &
\multicolumn{1}{|c|}{Avg Reqd Rank} & 4195 & 2031 & 376   \\ \cline{2-5}
\multicolumn{1}{|c|}{}                        &
\multicolumn{1}{|c|}{$95\%$ Interval Required Rank} & [3301,4985] & [1673,2553] & [309,459]     \\ \cline{2-5}
\multicolumn{1}{|c|}{}                        &
\multicolumn{1}{|c|}{Posterior Mean, $\theta_1$} & 0.496 & 0.352 & 0.105     \\ \cline{2-5}
\multicolumn{1}{|c|}{}                        &
\multicolumn{1}{|c|}{$95\%$ credible interval, $\theta_1$} & [0.087,0.993] & [0.085,0.761] & [0.042,0.289]     \\ \cline{2-5}
\multicolumn{1}{|c|}{}                        &
\multicolumn{1}{|c|}{ESS, $\theta_1$} & 85 & 119 & 147      \\ \cline{2-5}
\multicolumn{1}{|c|}{}                        &
\multicolumn{1}{|c|}{Posterior Mean, $\theta_2$} & 1.411 & 1.315 & 1.099     \\ \cline{2-5}
\multicolumn{1}{|c|}{}                        &
\multicolumn{1}{|c|}{$95\%$ credible interval, $\theta_2$} & [1.114,1.857] & [1.065,1.701] & [1.002,1.203]     \\ \cline{2-5}
\multicolumn{1}{|c|}{}                        &
\multicolumn{1}{|c|}{ESS, $\theta_2$} & 145 & 132 & 227     \\ \cline{2-5}
\multicolumn{1}{|c|}{}                        &
\multicolumn{1}{|c|}{Time} & 57213  & 53929 & 20869   \\ \cline{1-5}
\end{tabular}
}
\caption{Comparison of the different algorithms based on their performance in the experimental data sets in terms of preditive MSE and various posterior summaries for the unknown parameters. ESS stands for effective sample size.}
\label{t4}
\end{table}
\end{document}